\documentclass[aip,graphicx]{revtex4-1}

\draft 
\usepackage{amsmath}
\usepackage{graphicx,lineno} 
\usepackage[breaklinks,colorlinks=true,citecolor={blue}, linkcolor= {blue}, urlcolor ={blue}]{hyperref}

\begin{document}

\title{Torsional Sensitivity and Resonance Tuning in Width-Varying AFM Microcantilevers} 



\author{Le Tri Dat}
\email[]{letridat@dntu.edu.vn}
\affiliation{Engineering Research Group, Dong Nai Technology University, Bien Hoa City, Vietnam}
\affiliation{Faculty of Engineering, Dong Nai Technology University, Bien Hoa City, Vietnam}


\author{Nguyen Duy Vy}
\email[Corresponding author: ]{nguyenduyvy@vlu.edu.vn}
\affiliation{Laboratory of Applied Physics, Science and Technology Advanced Institute, Van Lang University, Ho Chi Minh City, Vietnam}
\affiliation{Faculty of Applied Technology, School of Technology, Van Lang University, Ho Chi Minh City, Vietnam}



\date{\today}

\begin{abstract}
The torsional vibration of atomic force microscope (AFM) cantilevers is key to high-resolution and high-sensitivity measurements. However, standard models often fail to accurately describe the dynamics of width-varying geometries. In this work, we present an exact analytical model for computing torsional resonance frequencies and mode shapes of overhang- and T-shaped microcantilevers. Our predictions match experimental torsional-to-flexural frequency ratios within 5\%, resolving long-standing discrepancies. We uncover the emergence of multiple spatial maxima in higher-order modes and demonstrate how overhang geometry allows tunable frequency shifts. Crucially, we derive a sensitivity function that quantifies the dependence of modal response on tip–surface coupling stiffness, revealing nontrivial geometry-dependent trends. These results offer clear design principles for enhancing AFM sensitivity via geometric control, providing a robust theoretical basis for optimizing next-generation microcantilever probes.world operational demands.
\end{abstract}


\maketitle 

\section{Introduction}\label{sec:intro}
Atomic force microscopy (AFM) has emerged as a powerful tool for high-resolution surface characterization and single-molecule force spectroscopy, offering exceptional sensitivity and versatility across diverse environments (e.g., air, liquid, and vacuum) \cite{NeumanNatMeth08, Sajjadi2017, Karimpour_micron_2017_vibration}. By monitoring cantilever deflections or shifts in resonant frequency, AFM enables the quantification of structural, thermal, and mechanical properties at the nanoscale \cite{Vinh2020APEX,Beytollah_JVC_2021_application, DatJJAP23}. Recent advances in cantilever dynamics have further expanded its applications, from biomolecular interaction mapping to nanomechanical property imaging \cite{15VyAPEX}.

While flexural vibration modes dominate conventional AFM operation, torsional modes have gained prominence for their unique advantages in specialized scenarios. For instance, torsional vibrations exhibit superior sensitivity for stiff materials and lateral stiffness measurements \cite{TurnerNanoTech01, 18HeinzeUltra}, enabling novel techniques such as sidewall probe imaging \cite{PayamMRT20}. Sharos et al. demonstrated that torsional modes achieve higher mass sensitivity than bending modes \cite{Sharos_apl_2004}, while Turner et al. highlighted their efficacy in higher-order mode applications \cite{TurnerNanoTech01}. Despite these benefits, existing models for torsional dynamics in width-varying cantilevers (e.g., overhang or T-shaped geometries) remain limited by approximations, leading to discrepancies between theory and experiments. This gap is particularly critical for next-generation AFM probes, where geometric tuning could unlock tailored frequency responses and enhanced sensitivity.

\begin{figure}[!ht] 
\includegraphics[width=0.8\textwidth]{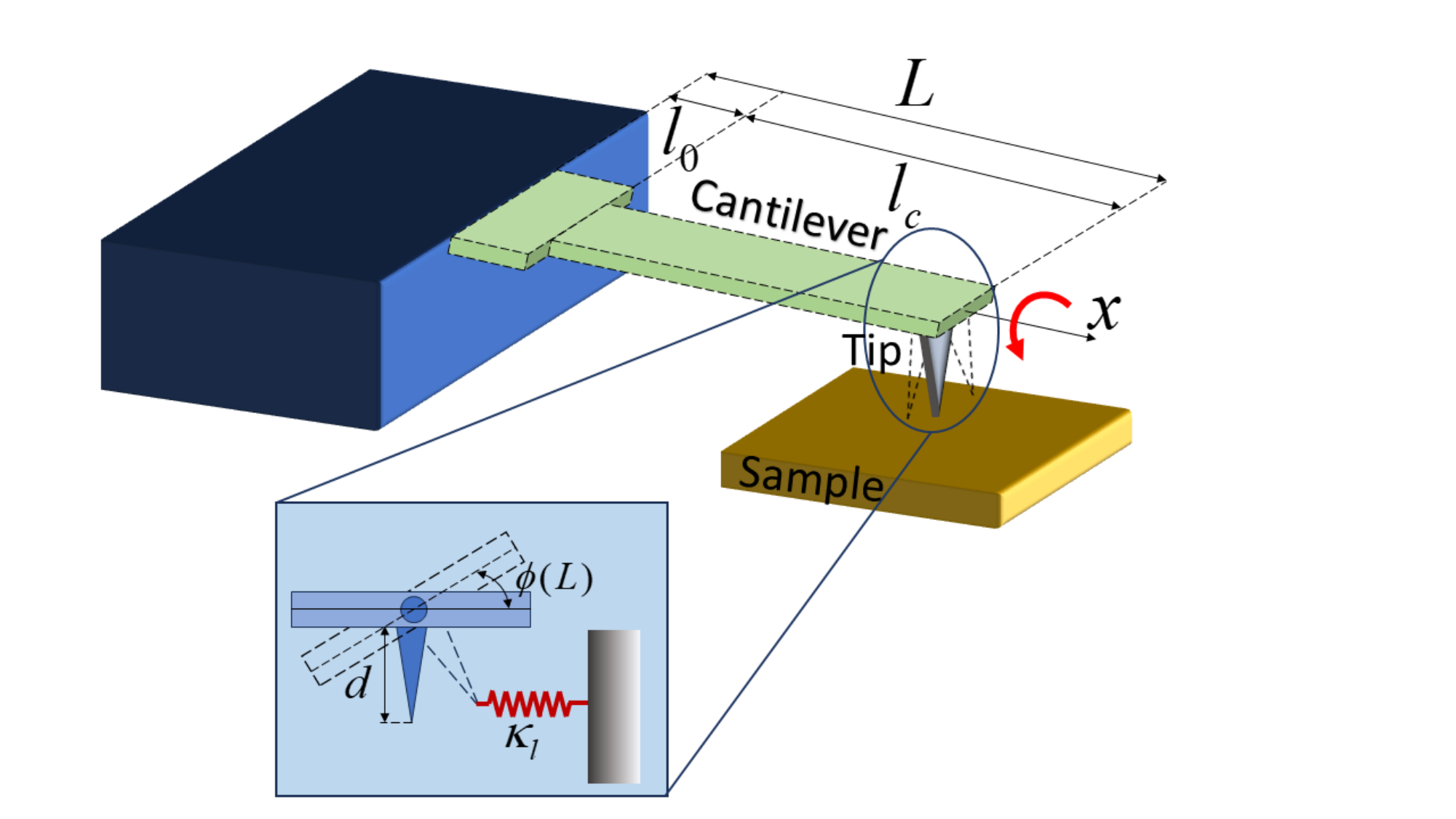}
\caption{A cantilever beam structure including an overhanging part of length $l_0$ and width $a_0$ is clamped at $x=0$ to the base (black region). For $a_0<a$, we have a T-shaped cantilever. The torsional modes are examined in the interaction with a sample via the effective interaction stiffness $\kappa_l$ (inset).} \label{fig:model}
\end{figure}

Beyond simple rectangular beams, complex geometries such as inverted T-shaped or V-shaped cantilevers \cite{PlazaAPL06,SaderRSI12} have attracted significant interest due to their enhanced performance. However, accurately determining their resonant frequencies and mode shapes remains challenging due to non-uniform width and thickness distributions. Zhang {\it et al.} \cite{ZhangJPD11} investigated variable-width cantilevers using polynomial approximations to circumvent analytical difficulties in solving the Euler-Bernoulli equation. While their approach revealed a geometric dependence of resonant frequencies, the solutions were computationally cumbersome and sensitive to polynomial assumptions. Plaza {\it et al.} \cite{PlazaAPL06} demonstrated that microcantilever arrays could mitigate initial deflections, yet precise frequency determination still relied heavily on experimental calibration. These limitations underscore the need for a robust analytical framework to predict the vibrational characteristics of non-uniform cantilevers—a gap this work addresses.

The overhang region, in particular, plays a critical role in cantilever dynamics, influencing both individual performance and coupled-array behavior. Recent studies show that the coupling strength between cantilevers in an array depends linearly on overhang length and inversely on the cubic power of overhang width \cite{DatMRT22}. This geometric sensitivity highlights the importance of accurate modeling, especially for applications requiring high-resolution surface topography or synchronized array measurements \cite{08APL_Chen_array,08array,23array}.

Our work bridges the gap between theory and experiment by providing precise analytical solutions for overhang-type cantilevers, reducing discrepancies in frequency predictions. Modifications to beam geometry, such as overhang length and width, can dramatically alter dynamic behavior, including the emergence of higher harmonic modes \cite{DatJJAP23}. While prior studies have explored various geometries—such as Payam {\it et al.} examining flexural spring constants in fluids \cite{Payam_APL_2018,VyIJSS22} and Plaza {\it et al.} optimizing T-shaped structures \cite{PlazaAPL06}—none have systematically resolved the torsional vibration challenges unique to overhang designs. This study fills that void, offering a unified approach to optimize cantilever performance for both single-probe and array-based applications.

Prior models fail to reconcile analytical predictions with experimental frequency ratios, as shown by Sadewasser et al. \cite{06RSI_Sadewasser}. Our work resolves this by 
presenting a comprehensive analytical and numerical investigation of how geometric modifications in overhang- and T-shaped cantilevers alter their torsional mode shapes, resonant frequencies, and—critically—their measurement sensitivity. Unlike prior studies focusing solely on frequency tuning, we systematically quantify the influence of overhang and T-section dimensions on both dynamic response and modal sensitivity, bridging a key gap in AFM cantilever design. Our analytical model predicts torsional-to-flexural frequency ratios within 3--5\% of Sadewasser et al.’s experimental data (solid blue circles), achieving an $R^2$ correlation coefficient of 0.98.

In Section~\ref{sec:analyt}, we derive exact frequency equations for T-shaped cantilevers (excluding external interactions) and establish how variations in overhang length, width, and T-junction geometry reshape mode shapes and harmonic spectra. 

Section~\ref{sec:result} presents validated results, highlighting two key advances:
(i) Geometric control of higher-order modes: We demonstrate how overhang tuning can selectively enhance or suppress specific harmonics, enabling tailored frequency responses.
(ii) Sensitivity-stiffness coupling: In Section~\ref{sec:sensit}, we introduce a generalized model for modal sensitivity incorporating sample interactions via the effective rigidity parameter, resolving prior discrepancies between theoretical and experimental sensitivity trends.

Our findings provide actionable guidelines for designing next-generation AFM probes, particularly for applications requiring high lateral resolution (e.g., nanoscale friction mapping) or synchronized array measurements. The conclusions in Section~\ref{sec:summ} summarize these insights and outline pathways for future experimental validation.

\section{Material and Methods}\label{sec:analyt}
The cantilever is assumed to be fabricated from silicon nitride \cite{19MohammadiUltra}, consistent with conventional AFM experiments, with a length 
$L$ = 200--500 $\mu$m, width $a$ = 35 $\mu$m, and thickness $t$ = 1.5 $\mu$m. 
Building on the analytical framework developed for flexural vibrations in nonuniform cantilevers \cite{DatMRT22,DatCiP20}, we extend this approach to model torsional vibrations in overhang- and T-shaped structures [see Fig.~\ref{fig:model}]. Our analysis systematically investigates how variations in the overhang and T-section dimensions influence the resonant frequencies and mode shapes of the cantilevers. Therefore, a detailed analysis on the multi-mode behavior of the cantilever is required. We found that, (i) higher-order torsional modes play a significant role in the vibration dynamics, necessitating a detailed analysis of their contributions,
and (ii) the dimensions of the overhang and T-parts critically affect both the frequency spectrum and mode shape distribution, highlighting the importance of precise geometric control for optimal performance.

The dynamic equation for the torsional vibration mode is written based on the Euler-Bernoulli theory of the beam \cite{meirovitch1967analytical, weaver1991vibration} as follows:
\begin{equation}\label{eq:torsion_eq}
    \frac{\partial}{\partial x}\left[ GJ(x)\frac{\partial \phi (x,t)}{\partial x}\right] -\rho I_{p}(x) \frac{\partial^2 \phi(x,t) }{\partial t^2} = 0, 
\end{equation}
where $\phi(x,t)$ is the deflection angle at position $x$ and time $t$. $G$ is the shear modulus and $\rho$ is the density of the beam. $J(x)$  and $I_p (x)$ are geometric functions of the beam cross section and the polar moment of inertia, respectively. Here, the cross section of the beam is a rectangular shape, hence, $J(x) = a(x)t^3 / 3$ and $I_p(x) = a^3(x)t/12$. It is shown that the width of the beam is $x$-dependent. Hence, the general solution of Eq. (\ref{eq:torsion_eq}) is $\phi(x,t)=\phi(x)e^{i\omega t}$. Substituting the general solution into Eq. (\ref{eq:torsion_eq}), one obtains the equations for the mode shape ($x$-dependent) and the natural frequency. The mode shape equation reads,
\begin{equation}\label{eq:torsion_notime_eq}
    \frac{d}{dx}\left[ GJ(x)\frac{d\phi (x)}{dx}\right] +\rho I_{p}(x)\omega^2\phi(x) = 0.
\end{equation}
For the current cross-section of the beam, the thickness of the overhang part and the outer cantilever part are assumed to be the same while the width is step-like with $x$, $a(x) = a_0$ if $x\leq l_0$ and $a(x) = a$ if $x> l_0$.
Equation (\ref{eq:torsion_notime_eq}) is divided into two equations. The first equation describes the overhang part,
$    \phi_0^{''}(x) + \gamma^2_0\phi_0(x) = 0, 
$
and the second equation is for the cantilever part,
${    \phi_c^{''}(x) + \gamma^2_c\phi_c(x) = 0.
}$
Here, $\gamma_{0,c} = \omega \sqrt{\frac{\rho I_{p,0,c}}{GJ_{0,c}}}$ is the characteristic frequency. Now, the frequency ratio is
$    \frac{\gamma_c}{\gamma_0} = \frac{a}{a_0}=\frac{1}{\kappa},
$ 
or, $\gamma = \gamma_c = \frac{1}{\kappa}\gamma_0$ could be used for brevity. 
The solutions of the differential equations 
are,
\begin{align}\label{eq:sol_overh}
    \phi_0(x) = A\sin(\kappa\gamma x) + B\cos(\kappa\gamma x), \\
    \phi_c(x) = C\sin(\gamma x)+D\cos(\gamma x). 
\end{align}
Solving the equation with suitable boundary conditions, 
we obtain a frequency equation,
\begin{equation}\label{eq:freq_eq}
    \kappa^2\cos(\gamma-\gamma\eta)\cos(\gamma\eta\kappa) -\sin(\gamma-\gamma\eta)\sin(\gamma\eta\kappa) = 0,
\end{equation}
where $\eta = l_0/L$; then the frequency is,
\begin{equation}\label{freq}
    \omega = \frac{\gamma}{L}\sqrt{\frac{GJ_c}{\rho I_{p,c}}}.
\end{equation}
And 
the mode shape is presented as,
    \begin{align}\label{mode shape}
        \phi_0(x) &= A\sin(\kappa\gamma x),\\
        \phi_c(x) &= 
        A\cos{\left[\left( 1-x\right)\gamma\right]}\sec{\left[\left( 1-\eta\right)\gamma\right]}\sin{\left( \kappa\eta\gamma \right)}
    \end{align}
The updated mode shapes, in the case of flexural vibration, have been shown to significantly modify that of the uniform cross-section cantilevers \cite{DatCiP20}. For the torsional modes, similar behavior is expected. 


\section{Results}\label{sec:result}
Typical microcantilever dimensions span the following ranges: length (
$L$) = 50--500 $\mu$m, width ($a$) = 10--50 $\mu$m, and thickness ($
t$) = 0.5--5 
$\mu$m. These ranges are consistent with experimental studies, such as those by \cite{Etayash2015} (silicon cantilevers: 
$L$ = 100--500 $\mu$m, $a$ = 20--50 
$\mu$m, $t$ = 0.3--2 $\mu$m and \cite{McFarland_2005} ($L\simeq$ 500 $\mu$m, $a$ = 97.2 $\mu$m, 
$t$ = 0.8 $\mu$m). In this work, we analyze silicon nitride microcantilevers (material properties in Table~\ref{tab:parameters}), focusing on how their mode shapes and resonant frequencies evolve with geometric modifications.

\begin{table}[!h]\centering
\caption{Parameters of cantilever part.}\label{tab:parameters}
\begin{tabular}{cccc}
 Parameters& Symbol (Unit) & Value  \\
\hline
Length & $L$ ($\mu$m)& 350& \\
Width& $a$ ($\mu$m)& 35& \\
Thickness& $t$ ($\mu$m)& 1.5&  \\
Young's modulus & $E$ (GPa)& 169& \\
Density & $\rho$ (kg/$m^3$)& 2300& \\
\end{tabular}
\end{table}
%

\begin{figure*}[!h] \centering 
\includegraphics[width=0.75\textwidth]{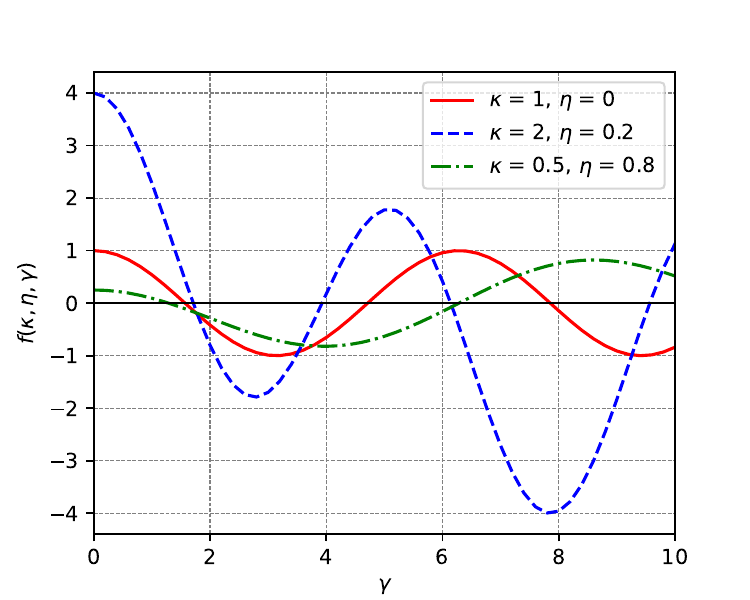} 
\caption{Equation \ref{eq:freq_eq} is illustrated for several geometric configurations characterized by the parameters $\kappa$ and $\eta$. The red solid line represents a rectangular cantilever with $\kappa = 1$ and $\eta = 0$, the blue dashed line corresponds to an overhang structure with $\kappa = 2$ and $\eta = 0.2$, and the green dash-dotted line denotes a T-shaped cantilever with $\kappa = 0.5$ and $\eta = 0.8$.
} \label{fig:freq_sols}
\end{figure*}

{
By solving Eq. (\ref{eq:freq_eq}), the resonant frequencies and corresponding mode shapes can be determined. Figure \ref{fig:freq_sols} illustrates how variations in the geometric parameters $\kappa$ and $\eta$ affect the oscillatory behavior of the solution along the horizontal axis. These oscillations correspond to the characteristic frequency solutions $\gamma$ associated with the vibration modes.

For the rectangular cantilever (represented by the solid red line), the first three modes occur at $\gamma = \pi/2$, $3\pi/2$, and $5\pi/2$, respectively. However, when the geometry is modified, these characteristic frequencies shift. For example, in the case of the overhang-shaped cantilever with $\kappa = 2$ and $\eta = 0.2$ (blue dashed line), the first mode appears at $\gamma \approx 1.712$, the second at approximately $5\pi/4$, and the third at $\gamma \approx 6.142$. In contrast, for the T-shaped cantilever with $\kappa = 0.5$ and $\eta = 0.8$ (green dash-dotted line), the first and second mode frequencies occur at $\gamma \approx 1.608$ and $\gamma \approx 7.854$, respectively.
}

\subsection{Changes in mode shapes}
\begin{figure*}[!h] \centering 
\includegraphics[width=0.9\textwidth]{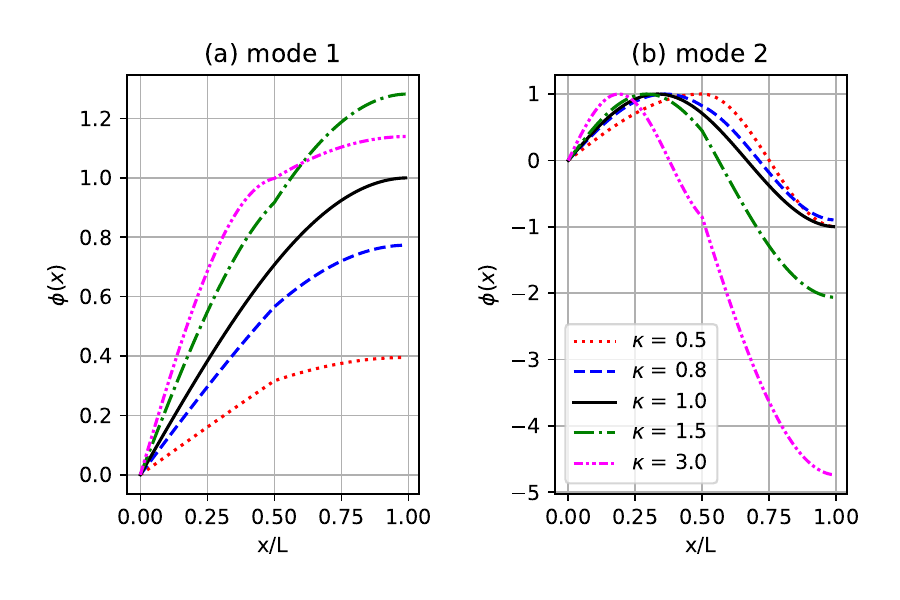} 
\caption{Mode shapes of cantilever beam for the two first modes with increasing the cantilever width via $\kappa$. Here, $\eta$ = 0.5 
. An increasing of $\kappa$ implies a wider cantilever.} \label{fig:mode_shape}
\end{figure*}
\begin{figure*}[h!] \centering
\includegraphics[width=\textwidth]{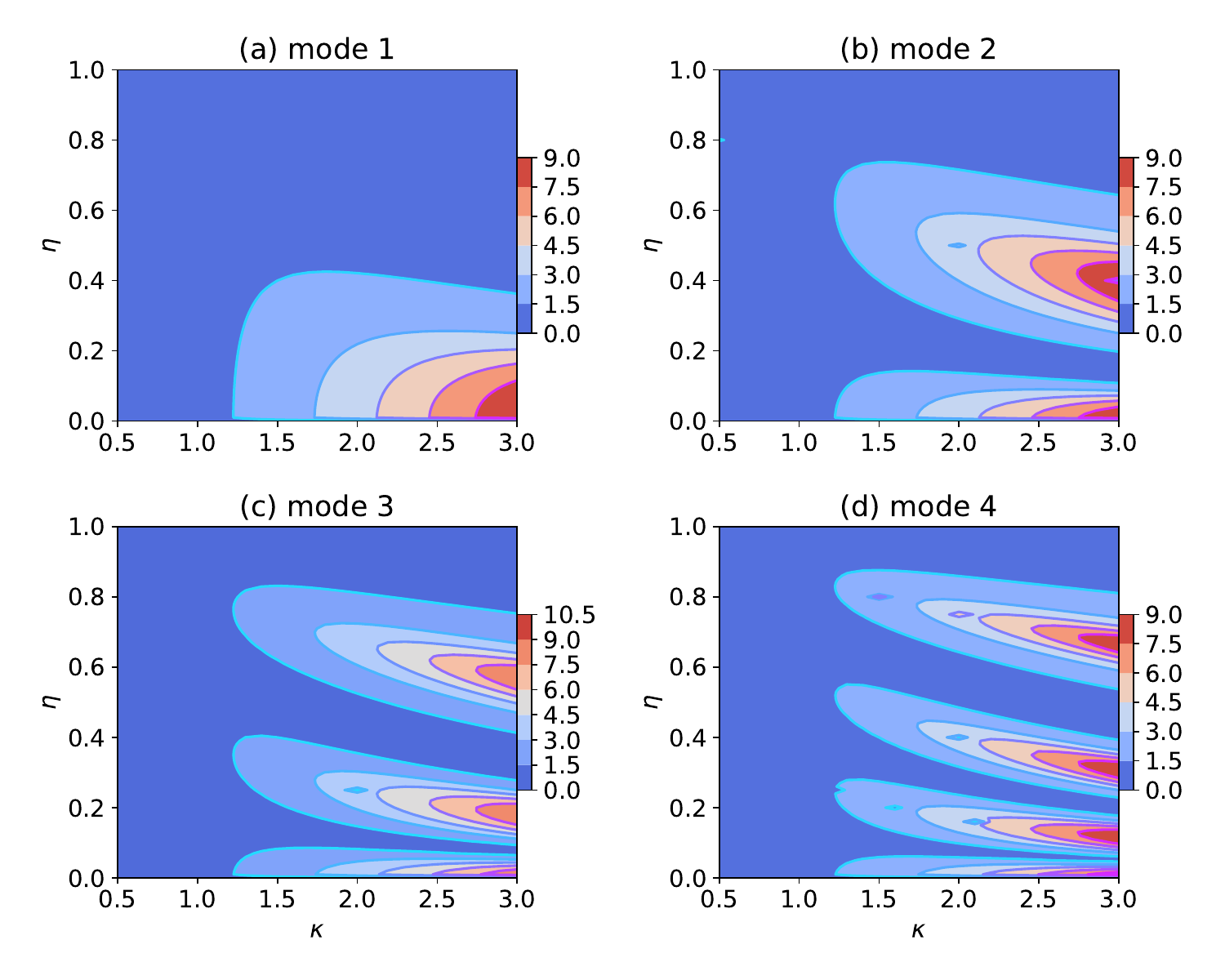} \vspace{-5pt}
\caption{The deflection angle at $L$, $x(L)/x(l_0)$, for the first four modes with several maxima. The number of maxima is proportional to the mode number and exists for $\kappa>1$.}\label{fig:phi_L}
\end{figure*}
The mode shapes have been expressed in Fig. \ref{fig:mode_shape} for (a) the first mode and (b) the second mode. Figure \ref{fig:mode_shape}(a) and \ref{fig:mode_shape}(b) are presented for $\eta=l_0/L = 0.5$ and various values of overhang widths, $\kappa=a_0/a$ = 1.0 by black-solid, 0.5 by red-dotted, 0.8 by blue dashed, 1.5 by green dash-dotted, and 3.0 by pink dash-dot-dotted lines. 
The deflection angle at $L$, $x(L)/x(l_0)$, has been shown to increase as the width of the cantilever increases. 

For the second mode, the mode shapes of $\kappa$ = 0.5--0.8 deviate from that of $\kappa = 1$ for 0 $<x<L$ with a maximum that then decreases and approaches the value -1 for $x=L$, while those of $\kappa>1$ greatly decrease (green dash-dotted lines and pink dash-dotted lines).

Figure~\ref{fig:phi_L} summarizes the behavior of the first four torsional modes, illustrating how the overhang affects the deflection angle at the beam's free end. The color intensity represents the deflection magnitude, revealing distinct maxima patterns for each mode:
\begin{itemize}
\item Mode 1: Single maximum (not explicitly located but observable in color gradient).
\item Mode 2: Two maxima at $l_0/L \simeq$ 0.05 and 0.4 (red regions).
\item Mode 3: Three maxima at $l_0/L \simeq$ 0.025, 0.2, and 0.6.
\item Mode 4: Four maxima at $l_0/L \simeq$ 0.01, 0.15, 0.3, and 0.65. 
\end{itemize}
The number of maxima consistently equals the mode number, demonstrating a clear correlation between modal order and spatial oscillation patterns. This systematic behavior confirms the predictable influence of the overhang geometry on torsional vibration characteristics.

\subsection{Changes in frequency}
\begin{figure*}[!ht] \centering
\includegraphics[width=1.2\textwidth]{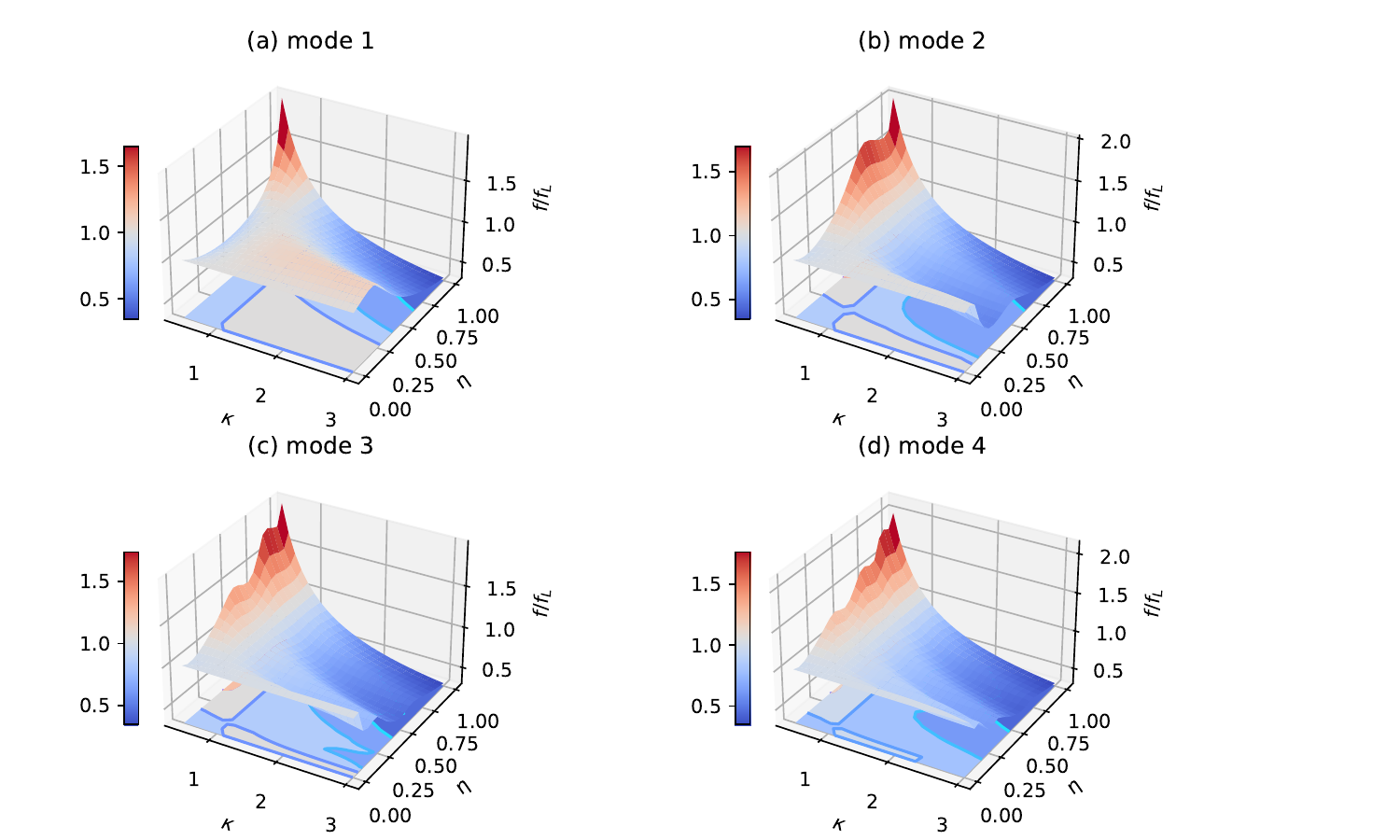}
\caption{The frequencies of the first four modes. (a) The first mode of the overhang-shaped ($\kappa>1$) presents a maximal frequency at 0 $\le\eta\le$ 0.25 while the T-shaped ($\kappa<1$) cantilevers have minima. (b)--(d) Higher modes have a tendency to reduce the frequency with $\eta$ and some small extrema appear.}\label{fig:freqs_wo_inter}
\end{figure*}

Examining the change in the cantilever frequency $f = \omega / 2\pi$, we observed an interesting nonmonotonic behavior in the dependence of $f$ on the overhang length $\eta$. The frequency of the first mode initially increases [Fig. \ref{fig:freqs_wo_inter}(a)], reaching a maximum at $\eta \simeq 0.2$, and then decreases rapidly as $\eta$ continues to increase. In the region near the maximum (from orange to red), the frequency increases with $\kappa$, i.e., with a larger overhang, which implies a stiffer cantilever.

The 2nd to 4th modes, on the other hand, no longer clearly exhibit a maximum. All frequencies tend to decrease rapidly as $\eta$ increases. For example, within the range $\eta = 0$--$0.5$, the second mode $f_2$ [Fig. \ref{fig:freqs_wo_inter}(b)] decreases significantly from 1.0 to $\simeq 0.5$ (1000 to 500 kHz). The 3rd and 4th modes [Fig. \ref{fig:freqs_wo_inter}(c) and (d)] display a small peak before dropping to lower frequencies.

These findings are notable for several reasons. First, there is a nearly flat plateau in the frequency response followed by a rapid decline. This feature could be advantageous for controlling and tuning cantilever frequencies by adjusting the length of the overhang section, thereby enhancing high-harmonic frequencies \cite{HoangJJAP17,DatJJAP23}.
Second, effective modulation of higher-order modes, governed by the structure's geometry, may enable the use of multiple modes in measurements, improve inter-mode coupling, and even support the appearance of high harmonics (since higher-order mode frequencies can be integer multiples of lower ones). Consequently, appropriate tuning of the overhang parameters can facilitate control over higher modes.

\begin{figure*}[!ht] \centering
\hspace{-0.5cm}\includegraphics[width=0.55\textwidth]{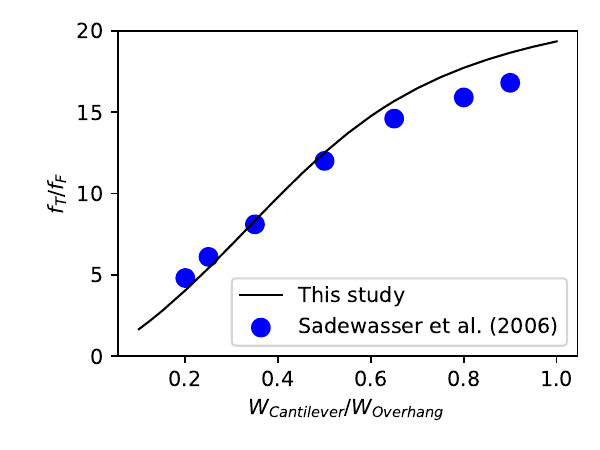}
\caption{Ratio of torsional to flexural frequencies of overhang-shaped cantilever from the experiment of Sadewasser et al. (solid blue circles) \cite{06RSI_Sadewasser} and this study (solid black line).  }\label{compared_w_exp}
\end{figure*}

Our approach also shows strong agreement with previous research on cantilever frequencies. For instance, Sadewasser et al. investigated the dynamics of cantilevers with an overhanging section \cite{06RSI_Sadewasser}. Compared to Sadewasser’s results, our method exhibits excellent consistency in the ratio between torsional ($f_T=\omega_T/2\pi$) and flexural ($f_F=\omega_F/2\pi$) frequencies under identical geometric conditions, as illustrated in Fig. \ref{compared_w_exp}.

\subsection{Torsional sensitivity}\label{sec:sensit}
The sensitivity of the flexural modes in overhung and T-shaped cantilevers has recently been studied \cite{VyIJSS22}, demonstrating that the dimensions of the overhang section can significantly influence the cantilever's frequency. For torsional modes, several investigations have been carried out by Abbasi {\it et al.} \cite{AbbasiMRT15}, focusing on cantilevers with sidewall probes in rectangular geometries. However, the torsional vibrations of overhung or T-shaped cantilevers have not yet been explored.

In this work, we analyze the torsional modal sensitivity of an overhung cantilever, assuming a tip-sample interaction modeled as a linear lateral spring with stiffness $\kappa_l$. This interaction is applied at the cantilever’s end position, $x = L$. The boundary condition at $x = L$ is expressed as ${\phi'(L) = -(\kappa_l d^2 / (GJ)) \phi(L) = -\beta_l \phi(L)}$, where $\beta_l = (\kappa_l d^2 / (GJ))$. This results in an additional term in the matrix $K$ in Eq. (\ref{eq:K_matrix}). The modified matrix $K$ is written as follows [Eq. (\ref{eq:K_matrix_w_interplay})],

    \begin{align}
    K_l = \left[\begin{matrix}
    	0 & 1 & 0 & 0\\
    	0 & 0 & \gamma\cos\gamma + \beta_l\sin\gamma& -\gamma\sin\gamma +\beta_l\cos\gamma \\
    	\sin\kappa\eta\gamma & \cos\kappa\eta\gamma & -\sin\eta\gamma & -\cos\eta\gamma \\
    	\kappa^2\cos\kappa\eta\gamma & -\kappa^2\sin\kappa\eta\gamma  & -\cos\eta\gamma & \sin\eta\gamma
    \end{matrix}\right].\label{eq:K_matrix_w_interplay}
    \end{align}

\begin{figure*}[!t]
\hspace{-0.5cm}\includegraphics[width=1.05\textwidth]{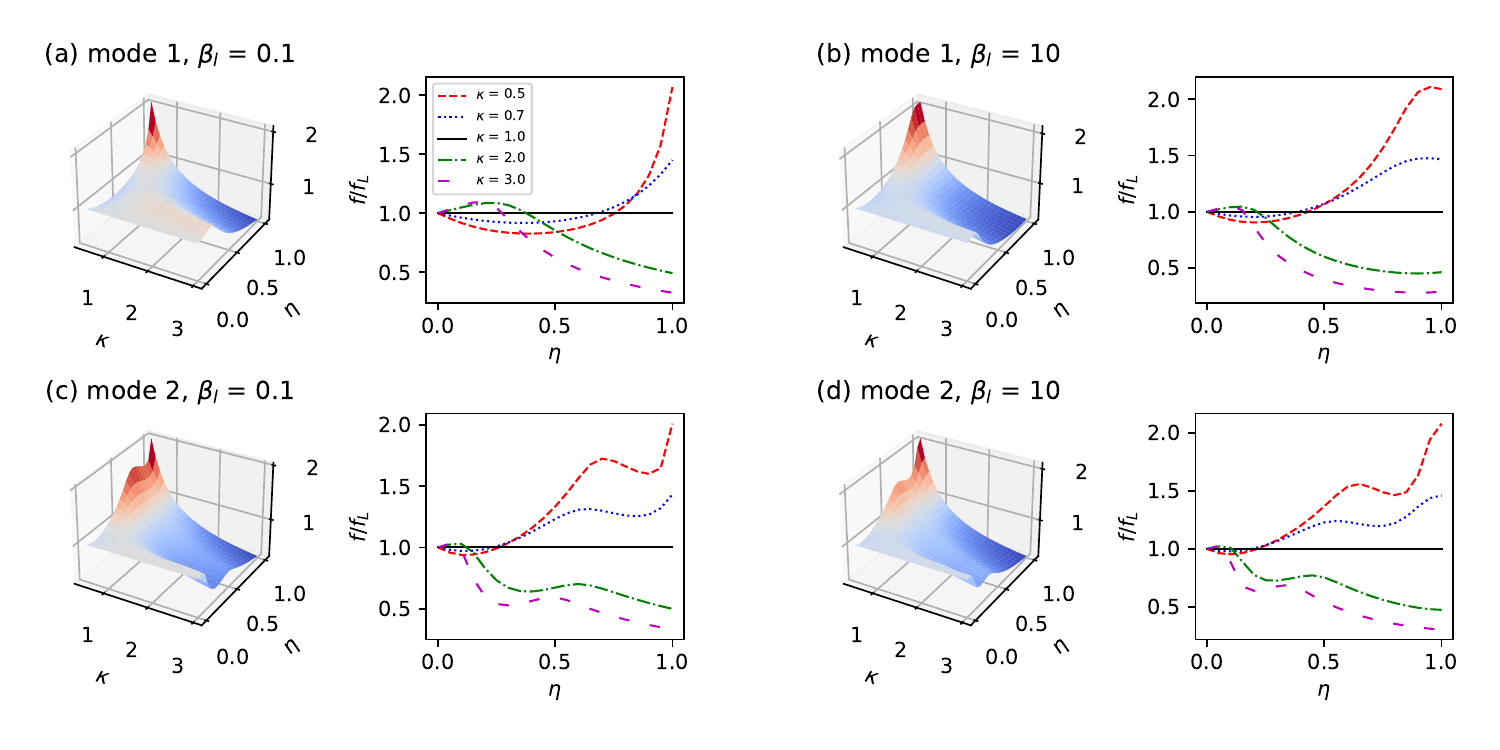}
\caption{(a) and (b) First torsional mode and (c) and (d) second torsional mode frequencies ($f$ normalized to the rectangular cantilever frequency ($f_L$ as functions of normalized overhang length ($\eta=l_0/L$ and coupling stiffness ($\beta_l$). Panels (a) and (c) show frequency trends across $\eta$ for varying 
$\kappa=w/a_0$, while (b,d) highlight cross-sections at specific 
$\kappa$ values (labeled). T-shaped cantilevers ($\kappa<$ 1, red dashed lines) exhibit frequency enhancement with increasing $\eta$, whereas overhang-shaped cantilevers ($\kappa>$ 1, violet/green lines) display frequency reduction. Critical balance points ($f/f_L$ indicate geometries where width-varying cantilevers match the torsional frequency of rectangular counterparts (
$\kappa$ = 1, black line). Non-monotonic shifts in $f$ with $\beta_l$ suggest tunable sensitivity via geometric and interaction parameter adjustments..  }\label{freq_2mode_w_inter}
\end{figure*}

Similarly, using the updated matrix $K_l$, a characteristic equation was obtained by calculating the determinant $C$ of the matrix. Finally, a characteristic equation for the frequency is obtained,
\begin{align}\label{eq:freq_eq_w_interplay}
 C=& C_0(\kappa, \eta, \gamma) + C_{int}(\kappa, \eta, \beta_l,\gamma),
\end{align}  
where
    \begin{align}
   C_0=& \gamma\{ \kappa^2 \cos\left[ \gamma(1 - \eta)\right]\cos(\gamma\eta\kappa)-\sin\left[ \gamma(1 - \eta) \right]\sin(\gamma\eta\kappa) \},\\
    C_{int}= & \beta_l \{ \kappa^2 \sin\left[ \gamma(1-\eta) \right]\cos(\gamma\eta\kappa) + \cos(\gamma(1-\eta)\sin(\gamma\eta\kappa) \}.
    \end{align}
$C_0$ and $C_{\text{int}}$ represent the non-contact and contact contributions, respectively. The variation of frequency with respect to $\eta$ and the coupling stiffness $\beta$ is shown in Fig. \ref{freq_2mode_w_inter} for the first and second modes. It is observed that, for $\kappa < 1$, the frequency $f$ first decreases and then increases with $\eta$, whereas for $\kappa > 1$, the trend is reversed. Notably, there exists a balance point at which $f / f_L = 1$ for every value of $\kappa$. This is particularly interesting because one can select the length and width of an overhung or T-shaped cantilever such that it yields the same torsional frequency as a rectangular one. In other words, these points could serve as a reference for designing width-varying cantilevers.

Increasing the coupling stiffness $\beta$ causes the frequency to change more rapidly. For example, in the first mode [see Fig. \ref{freq_2mode_w_inter}(b)], the curve intersects the $f/f_L = 1$ line earlier and then either increases (for $\kappa < 1$) to reach a maximum (as shown by the red dashed and blue dotted lines), or decreases (for $\kappa > 1$) to reach a minimum (violet long-dashed and green dash-dotted lines). For higher modes, additional extrema may appear, following a trend similar to that shown in Fig. \ref{fig:freqs_wo_inter}.

The sensitivity is defined as the change in frequency with respect to the interaction strength \cite{TurnerNanoTech01},
\begin{align}\label{eq:sensivity}
    S =\frac{\partial \omega}{\partial\beta_l} = \frac{\partial\omega}{\partial\gamma}\frac{\partial\gamma}{\partial\beta_l}
    =
    \frac{\partial\omega}{\partial\gamma}\left(-\frac{\partial C / \partial\beta_l}{\partial C / \partial\gamma}\right).
\end{align}

Here, the frequency of the beam is computed by
\begin{equation}\label{eq:freq_sen}
    \omega = \gamma\sqrt{\frac{GJ}{\rho I_p}}.
\end{equation}
Then, the normalized torsional sensitivity is obtained.
\begin{align}\label{eq:full_sensit}
    \sigma_T = \frac{\kappa^2 \sin\left[ \gamma(1-\eta)\right]\cos(\gamma\eta\kappa) + \cos\left[ \gamma (1 - \eta) \right]\sin(\gamma\eta\kappa)}{D},
\end{align} where

\begin{widetext}
    \begin{align}\label{eq:denom}
    D =& \cos\left[ \gamma (1 - \eta) \right] \left\{ \kappa\left[ -\beta_l\eta + (-1 + \beta_l (-1 + \eta))\kappa \right]\cos(\gamma\eta\kappa) + 
    \gamma\left[ 1 + \eta(-1 + \kappa^3) \right]\sin(\gamma\eta\kappa) \right\}
    +\nonumber\\
    &+ \sin\left[ \gamma (1 - \eta) \right]\left\{ \gamma\kappa\left[ \eta + \kappa - \eta\kappa\right]\cos(\gamma\eta\kappa) 
    +
    \left[ 1 + \beta_l + \beta_l\eta ( -1 + \kappa^3) \right]  \right\}\sin(\gamma\eta\kappa).
    \end{align}
\end{widetext}

\begin{figure*}[!h] \includegraphics[width=1.05\textwidth]{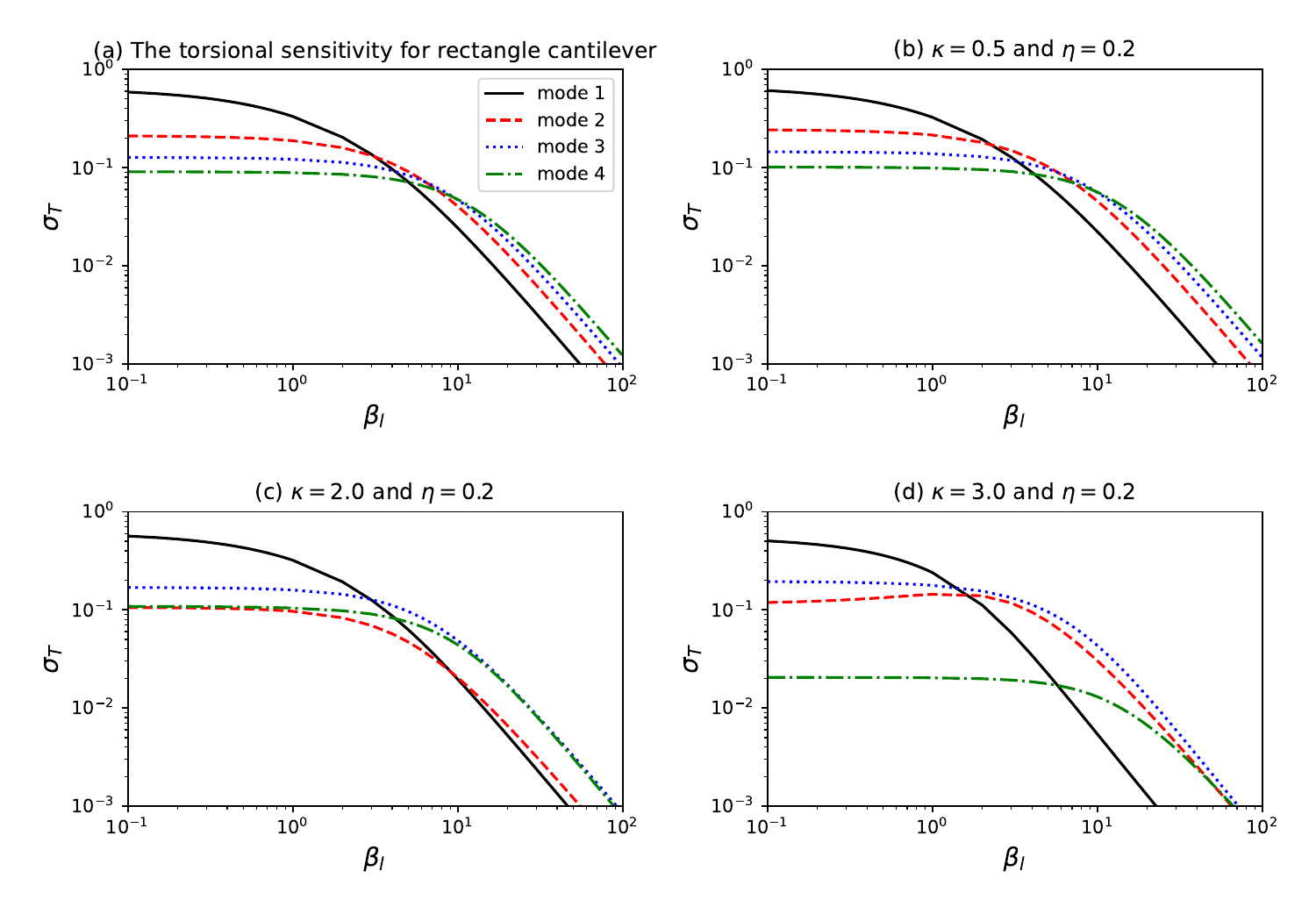} \vspace{-.5cm}
\caption{Sensitivity $\sigma_T$ of the first four torsional modes for various value of $\kappa$. (a) $\kappa$ = 1 (rectangular cantilever). (b) $\kappa$ = 0.5 (T-shaped cantilever). (c) and (d) $\kappa>$ 1 (overhang-shaped cantilevers). For the overhang-shaped cantilevers, $\sigma_T$ reduces with $\beta_l$ faster for wider overhang width $\kappa$. Here, $\eta=0.2$ is used.}\label{fig:sensitivity_1}
\end{figure*}
The normalized torsional modal sensitivity is shown in Fig. \ref{fig:sensitivity_1}: (a) $\kappa=1$ for the rectangular, (b) $\kappa=0.5$ for the T-shaped, (c) $\kappa=2.0$ for the overhang-shaped, and (d) $\kappa=3$ for the wider overhang-shaped cantilevers. Here, $\eta=0.2$ is used. 
It is recognized that the formulation of the sensitivity of a rectangle cantilever has been obtained if the geometric ratios were set $\kappa = 1$ or $\eta = 0$, and the analytical calculation of Turner {\it et al.} is realized \cite{TurnerNanoTech01}.

First, a similar behavior in the modal sensitivity ($\sigma_T$) of a rectangular cantilever, as reported by Turner {\it et al.} \cite{TurnerNanoTech01}, is reproduced in Fig. \ref{fig:sensitivity_1}(a), with the note that $\sigma_T$ values here are slightly higher due to the use of a longer cantilever length of $L = 350\ \mu$m [Turner used a 200\ $\mu$m-long cantilever].

For T-shaped cantilevers [Fig. \ref{fig:sensitivity_1}(b)] $\sigma_T$ is slightly modified, whereas for overhang-shaped cantilevers [Fig. \ref{fig:sensitivity_1}(c)--(d)] it is significantly altered. The $\sigma_T$ of the first mode (black solid line) decreases more rapidly than those of the higher modes, while the $\sigma_T$ of the fourth mode remains relatively stable over a wide range of $\beta_l$ (green dash-dotted line). Notably, at low $\beta_l$, the third mode (blue dotted line) shows increased sensitivity, surpassing that of the second mode—an inversion compared to the conventional sensitivity order $\sigma_T^{\text{mode 1}} > \sigma_T^{\text{mode 2}} > \sigma_T^{\text{mode 3}} > \sigma_T^{\text{mode 4}}$ seen in rectangular and T-shaped cantilevers.

Furthermore, an increase in sensitivity is observed only for the second mode within the range $\beta_l = 1$--10. This differs from the typical trend, where sensitivity tends to increase over a wider range, typically $\beta_l = 1$--100, depending on the mode number, before dropping rapidly to zero as $\beta_l \rightarrow 10^3$, as shown in Ref. \cite{TurnerNanoTech01} for torsional modes and in Refs. \cite{PayamUltra13, VyIJSS22} for flexural modes.

To further enhance the dynamic response and sensitivity of torsional modes, cantilevers equipped with extended sidewall probes could be employed, as demonstrated in Ref. \cite{DaiMST07}. These have shown interesting behavior depending on the geometry of the extended probe \cite{PayamMRT20}. We intend to explore this topic in a future study.

\section{Conclusions}\label{sec:summ}
In this study, we analytically derived the frequency characteristic equations and modal sensitivities for the torsional modes of overhang- and T-shaped cantilevers. Our results reveal significant and effective changes in both mode shape and frequency as functions of the overhang length, providing a versatile framework for selecting dimensional parameters to achieve specific performance characteristics. Notably, we presented a detailed analysis of the modal sensitivity of these cantilevers for the first time.

By modeling the tip–sample interaction as a linear lateral spring, we introduced a tunable coupling stiffness parameter $\beta_l$ that directly affects the torsional response. We showed that increasing $\beta_l$ leads to a non-monotonic shift in the modal frequencies, with critical balance points where the torsional frequency of width-varying cantilevers matches that of rectangular ones—offering practical benchmarks for cantilever design. The influence of $\beta_l$ on modal sensitivity was found to be strongly geometry-dependent, particularly for overhang-shaped cantilevers where unconventional sensitivity ordering emerged. This behavior suggests that geometric modifications can be strategically used to enhance or suppress sensitivity for specific modes.

The close match with Sadewasser’s experiments—a widely cited benchmark for AFM cantilever dynamics—validates our model’s ability to capture real-world torsional-flexural coupling without relying on empirical fitting. Minor deviations ($<$ 5\%) may arise from differences in material properties (e.g., silicon vs. silicon nitride) or edge effects in Sadewasser’s fabrication process.

Our analytical findings not only align with previous studies for rectangular cantilevers \cite{TurnerNanoTech01}, but also extend the analysis to geometries that had not been previously explored in the context of torsional modes. These insights offer valuable guidance for experimentalists in designing cantilever structures with tailored frequencies and sensitivities, enabling highly sensitive measurements across a range of applications. Future work may consider incorporating extended sidewall probes, which have shown promising results in improving dynamic behavior and sensitivity \cite{DaiMST07, PayamMRT20}.

\section*{Acknowledgements}
We thank Amir F. Payam (Ulster University) for fruitful discussion and encouragement.
   
\bibliography{jjap_modal_sens}

\end{document}